\def\lesssim{\mathrel{\hbox{\rlap{\hbox{\lower4pt\hbox{$\sim$}}}\hbox{$<$}}}}

\def\gtrsim{\mathrel{\hbox{\rlap{\hbox{\lower4pt\hbox{$\sim$}}}\hbox{$>$}}}}

\def\msun{M$_{\odot}$}

\def\ll_lsun{Log$({L/\rm L_{\odot}})$~}

\def\masa_msun{$M/ \rm M_{\odot}$~}

\def\m_mstar{$M/M_{*}$~}

\documentclass{aa}
\usepackage{graphicx}
        
\begin{document}

\title{New DA white dwarf evolutionary models and their pulsational
properties}

\author{A. H. C\'orsico,\thanks{Fellow of the Consejo Nacional de 
Investigaciones Cient\'{\i}ficas y T\'ecnicas (CONICET), Argentina.}
L. G. Althaus,\thanks{Member of the Carrera del Investigador
Cient\'{\i}fico y Tecnol\'ogico, CONICET, Argentina.} 
O. G. Benvenuto\thanks{Member  of  the  Carrera  del Investigador
Cient\'{\i}fico, CIC, Argentina.}
\and A. M. Serenelli \thanks{Fellow of CONICET, Argentina.} }

\offprints{A. H. C\'orsico}

\institute{Facultad  de  Ciencias
Astron\'omicas  y Geof\'{\i}sicas, Universidad  Nacional de  La Plata,
Paseo del Bosque S/N, (1900) La Plata, Argentina\\
\email{acorsico,althaus,obenvenu,serenell@fcaglp.fcaglp.unlp.edu.ar} }

\date{Received; accepted}

\abstract{In this letter we  investigate the pulsational properties of
ZZ  Ceti stars on  the basis  of new  white dwarf  evolutionary models
calculated  in a  self-consistent  way with  the  predictions of  time
dependent element  diffusion and  nuclear burning.  In  addition, full
account is taken  of the evolutionary stages prior  to the white dwarf
formation.   Emphasis is  placed on  the trapping  properties  of such
models. By  means of adiabatic, non-radial  pulsation calculations, we
find,  as a result  of time  dependent diffusion,  a much  weaker mode
trapping  effect,  particularly  for  the high-period  regime  of  the
pulsation g-spectrum.  This result is  valid at least for  models with
massive hydrogen-rich  envelopes. Thus, mode trapping would  not be an
effective  mechanism to  explain the  fact that  all the  high periods
expected  from standard  models  of stratified  white  dwarfs are  not
observed in the ZZ Ceti stars.
 }  

\authorrunning{C\'orsico et al.}

\titlerunning{New DA white dwarf evolutionary models and their 
pulsational properties}

\maketitle


\section{Introduction}

Pulsating  DA white dwarfs  (WD) or  ZZ Ceti  stars have  captured the
attention of  numerous researchers  since the first  star (HL  Tau 76,
Landolt  1968)  belonging  to  this  class  was  reported  to  exhibit
multi-periodic luminosity variations (McGraw 1979).  Over the last two
decades, various studies have presented strong evidence that pulsating
DA WDs represent  an evolutionary stage in the  cooling history of the
majority, if  not all, DA  WDs. Rapid progress  in the study  of these
pulsating  stars  has  been  possible  thanks to  the  development  of
powerful  theoretical  tools paralleled  by  an  increasing degree  of
sophistication in observational techniques.   A major step towards the
understanding of  ZZ Ceti  pulsations was given  by Dolez  \& Vauclair
(1981) and Winget  et al.  (1982) who  independently demonstrated that
models of ZZ Ceti stars have pulsationally unstables $g-$modes due to the 
$\kappa-\gamma$ mechanism.
From then on,  the asteroseismology of DA WDs  has provided invaluable
insights on  their internal structure and evolution.   (Tassoul et al.
1990; Brassard et al.  1991,  1992ab; Gautschy et al. 1996 and Bradley
1996, 1998, 2001 amongst others).

An  important aspect  of  pulsating  WDs is  related  to the  trapping
properties.  Mode trapping in  compositionally stratified WDs has been
invoked to explain  the longstanding fact that all  the modes expected
from theoretical models are not actually observed in the ZZ Ceti stars
(Winget  et al.   1981; Brassard  et al.   1992a).  In  this scenario,
certain  modes  are  characterized   by  local  wavelengths  that  are
comparable  to  the  thickness  of  one of  the  compositional  layer,
particularly  the hydrogen-rich envelope.   The importance  of trapped
modes lies on  the fact that they  appear to be the most  likely to be
observed because they require low kinetic energies to reach observable
amplitudes.  More specifically, the amplitude of the eigenfunctions of
modes trapped  in the  hydrogen envelope is  small in the  core, which
causes such modes  to have low oscillation kinetic  energy as compared
with adjacent modes.  This  behaviour manifests itself as local minima
in kinetic energy versus period diagrams. In particular, trapped modes
characterized  by  periods close  to  the  thermal  time-scale of  the
driving  region will  reach  high  enough amplitudes  for  them to  be
observed.    This  picture  has   been  reinforced   by  non-adiabatic
calculations (Dolez \& Vauclair  1981; Winget et al.  1982).  However,
recent evidence seems  to cast some doubts on  the correlation between
observed amplitudes  and mode trapping.   Indeed, recent seismological
studies of ZZ  Ceti stars (e.g.  Bradley 1998) point  to the fact that
the  observed  periods having  the  largest  amplitudes  in the  power
spectrum do not  correspond to trapped modes as  predicted by the best
fitting model.

The pulsation  properties depend  on the details  of the  WD modeling.
This is particularly true  regarding the abundance distribution at the
chemical  interfaces, mostly  at the  hydrogen-helium  transition.  In
this connection,  most of  the existing calculations  invoke diffusive
equilibrium in the trace element  approximation to assess the shape of
the hydrogen-helium transition (Tassoul  et al.  1990; Brassard et al.
1992ab;  Bradley 1996).   However, equilibrium  conditions may  not be
achieved  at the  base  of  massive hydrogen  envelopes,  even at  the
characteristic ages of ZZ Ceti stars (see Iben \& MacDonald 1985).  In
view of these concerns, we  have recently carried out new evolutionary
calculations  for DA  WD  stars  which take  fully  into account  time
dependent element diffusion, nuclear burning and the history of the WD
progenitor in a  self-consistent way.  The present letter  is aimed at
specifically exploring the trapping properties of such models.


\section{Evolutionary models}

The  WD  models on  which  the present  results  are  based have  been
calculated by means of a detailed evolutionary code developed by us at
La Plata Observatory.  The code has been employed  in previous studies
on  WD evolution  (Althaus et  al. 2001ab)  and it  has  recently been
modified to  study the evolutionary  stages prior to the  WD formation
(see  Althaus  et  al.   2001c).  The  constitutive  physics  include:
up-to-date  OPAL  radiative  opacities  for  different  metallicities,
conductive opacities, neutrino emission  rates, a detailed equation of
state  and a  complete  network of  thermonuclear  reaction rates  for
hydrogen and helium  burning (see Althaus et al.  2001c). For a proper
treatment  of   the  diffusively  evolving   chemical  stratification,
gravitational  settling  and the  thermal  and  chemical diffusion  of
nuclear species have been considered.

The  evolutionary stages  prior to  the WD  formation have  been fully
taken into account. Specifically, we started our calculations from a 3
\msun~ star at the zero-age main sequence and we follow its further
evolution all the way from the stage of hydrogen and helium burning in
the core  up to the  tip of the  asymptotic giant branch  where helium
thermal pulses occur.  After experiencing 11 thermal pulses, the model
is forced  to evolve towards  its WD configuration by  invoking strong
mass loss episodes.  As a result, a WD remnant of 0.563
\msun~ is obtained. The evolution of this remnant is pursued through the 
stage of planetary nebulae nucleus to the domain of the ZZ Ceti stars
on the WD  cooling branch.  An important aspect  of these calculations
is related  to the evolution of  the chemical abundance  during the WD
cooling. In particular, the  shape of the composition transition zones
is of  the utmost importance  regarding the pulsational  properties of
the ZZ Ceti  models.  In this respect, diffusion  processes cause near
discontinuities  in the  abundance distribution  at the  start  of the
cooling branch to be considerably smoothed out by the time the ZZ Ceti
domain  is reached.  This  can be  appreciated in  Fig. 1,  which also
illustrates  the profile  of the  hydrogen-helium  interface resulting
from  the predictions of  diffusive equilibrium  in the  trace element
approximation (thin  dotted line). The  shape of the  innermost carbon
and  oxygen distribution  emerges from  the  chemical rehomogenization
process  due to  the  Rayleigh-Taylor instability  occurring at  early
stages of the WD evolution (see Althaus et al.  2001c and also Salaris
et  al.   1997)\footnote{Before  rehomogenization,  the shape  of  the
carbon and  oxygen profile towards  the centre is characterized  by an
off-centered  peak  typical  of  evolutionary  calculations  in  which
semi-convection and overshooting are not considered (see Mazzitelli
\& D'Antona 1986).}. Surrounding the carbon-oxygen interior there is
a shell  rich in  both carbon ($\approx$  35\%) and  helium ($\approx$
60\%), and a overlying layer  consisting of nearly pure helium of mass
0.003 \msun.  The  presence of carbon in the  helium-rich region below
the helium buffer stems from the short-lived convective mixing episode
that has  driven the carbon-rich zone  upwards during the  peak of the
last helium pulse on the  asymptotic giant branch.  We want to mention
that  the total  helium  content  within the  star  once helium  shell
burning is  eventually extinguished amounts  to 0.014 \msun~  and that
the mass of  hydrogen that is left at the start  of the cooling branch
is about  $ 1.5 \times 10^{-4}$  \msun, which is reduced  to $7 \times
10^{-5}$ \msun~ due  to the interplay of residual  nuclear burning and
element diffusion by the time the ZZ Ceti domain is reached.

\section{Results}

Next, we  shall discuss  the pulsational properties  of a  selected WD
model at $T_{\rm  eff} \sim 12000$ K. We  should remark that, although
the  chemical  profiles  evolve  as  the WD  cools  down  through  the
instability strip (see Althaus et  al.  2001c), the conclusions of the
present  paper remain valid  for any  model belonging  to the  ZZ Ceti
instability strip.   We begin by showing  in Fig. 2 the  square of the
Brunt-V\"ais\"al\"a  frequency   $N$  (computed  as   in  Brassard  et
al. 1991) and the Ledoux term $B$ of such a model. The results for the
diffusive   equilibrium  approximation  are   also  plotted   as  thin
lines. Note the smooth  shape of $B$, which is a direct consequence
of  the chemical abundance  distribution.  The  contributions
from  the  Ledoux  term  are  characterized  by  extended  tails,  and
translate into smooth bumps on $N^2$.

The  characteristic of $B$  and $N^2$  as predicted  by our  models is
markedly different from  those found in previous studies  in which the
WD evolution  is treated in  a simplified way,  particularly regarding
the  chemical  abundance distribution  (e.g.   Tassoul  et al.   1990;
Brassard   et   al.     1991,   1992ab;   Bradley   1996).    Clearly,
non-equilibrium  chemical profiles  lead to  $B$ values  with markedly
less  pronounced  peaks as  compared  with  the diffusive  equilibrium
treatment. Accordingly, the Brunt-V\"ais\"al\"a frequency turns out to
be smoother as a result of non-equilibrium diffusion.

For the pulsation analysis we have employed the general Newton-Raphson
code  described in C\'orsico  \& Benvenuto  (2001).  We  have computed
g-modes with  $\ell= 1, 2$  and 3  with periods in  the range of  50 s
$\lesssim P_k \lesssim $ 1300 s ($k$ being the radial order of modes).
The upper panels of Figs. 3, 4 and 5 show, respectively, the values of
oscillation kinetic energy for modes with  $\ell= 1, 2$ and 3 in terms
of computed periods.  Lower panels depict the corresponding values for
the  forward period  spacing $\Delta  P_k$ ($\equiv  P_{k+1}  - P_k$).
Filled  dots  depict  the  results  corresponding to  our  model  with
non-equilibrium  diffusion, whereas  empty dots  indicate  the results
predicted   by  the  diffusive   equilibrium  approximation   for  the
hydrogen-helium interface.   In the  interests of clarity,  the scale
for  the kinetic  energy  in  the case  of  diffusive equilibrium  is
displaced upwards by 1 dex.

For  the  non-equilibrium   diffusion  model  the  quantities  plotted
(especially the  $E_{\rm kin}$  values) exhibit two  clearly different
trends.  Indeed, for $P_k \gtrsim 600$ s and irrespective of the value
of  $\ell$, the  kinetic energy  of adjacent  modes is  quite similar,
which  is in  contrast with  the  situation found  for lower  periods.
Interestingly, the  $\Delta P_k$ minima are commonly associated with  
$E_{\rm kin}$ maxima, but that modes with $E_{\rm kin}$ maxima are
adjacent to the modes with $E_{\rm kin}$ minima.
On the  other hand, the period spacing diagrams
show appreciable variations of $\Delta  P_k$ for $P_k \lesssim 600$ s.
This is due  mostly to the presence of  chemical abundance transitions
in DA WD models as explained by Brassard et al (1992ab).  In contrast,
for higher periods  the $\Delta P_k$ of the modes  tend to a constant,
asymptotic value (Tassoul 1980).

The  assumption   of  diffusive  equilibrium  in   the  trace  element
approximation in WD  modeling gives rise to a  kinetic energy spectrum
and  period spacing  distribution in  which the  presence of  the well
known  mode  trapping phenomenon  is  clearly  visible, as  previously
reported  by  numerous  investigators  (see  Brassard  et  al.  1992b,
particularly their  figures 20a and 21a  for the case of  $M_{\rm H} =
10^{-4}  M_*$).  The  trapped  modes correspond  to  modes with  local
minima in $\log (E_{\rm kin })$ and $\Delta P_{k}$.  Here, we find
that these  trapping properties virtually vanish when  account is made
of  WD  models with  diffusively  evolving  stratifications.  This  is
particularly true  for large periods, though for  low periods trapping
is  also substantially  affected. We  attribute the  differences found
between both treatments to the markedly different shapes of the Ledoux
term  at  the  hydrogen-helium  interface  as predicted  by  non-  and
equilibrium diffusion.

From  the results presented  in this  letter we  judge that,  for high
periods, trapping mechanism in  massive envelopes of stratified WDs is
not an appropriate one to explain the fact that all the modes expected
from  theoretical models are  not observed  in ZZ  Ceti stars.   It is
worth mentioning that Gautschy  \& Althaus (2001) have recently found,
on the  basis of  a consistent diffusion  modeling, a  weaker trapping
effect  on the  periodicities  in  DB WDs.   Our  results give  strong
theoretical support to recent evidence against the claimed correlation
between the  observed luminosity variations amplitude  and trapping of
modes.   Finally, to  place  these  assertions on  a  firmer basis,  a
non-adiabatic  stability  analysis of  the  pulsational properties  of
non-equilibrium  diffusion  models  is  required.   A  more  extensive
exploration of the results presented  in this letter will be presented
in a future work.

\begin{acknowledgements}

We warmly acknowledge to our referee, Paul Bradley, for the effort 
he invested in the revision of our article. His comments and suggestions
strongly improved the original version of this work.  

\end{acknowledgements}

\newpage

\begin{figure*}
\centering
\includegraphics[width=500pt]{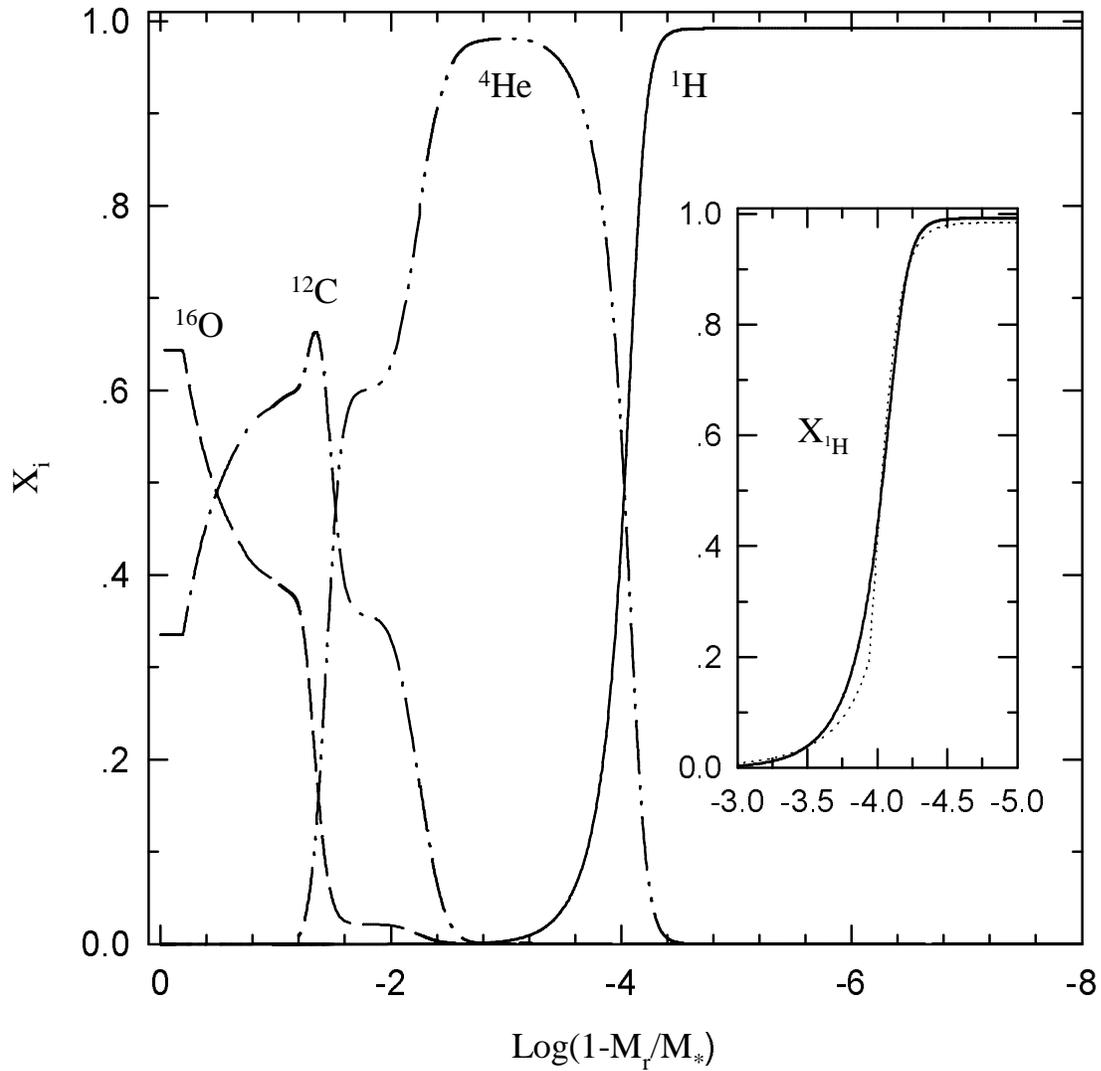}
\caption{Internal chemical profiles for hydrogen, 
helium,  carbon and  oxygen. The  hydrogen-helium  interface resulting
from   the   predictions   of  non-equilibrium   diffusion   (diffusive
equilibrium) are shown  with solid line (thin dotted line) in the inset. 
The WD stellar mass is $0.563 {\rm M_{\odot}}$ and the effective
temperature is $\approx$ 12000K.}
\end{figure*}

\begin{figure*}
\centering
\includegraphics[width=500pt]{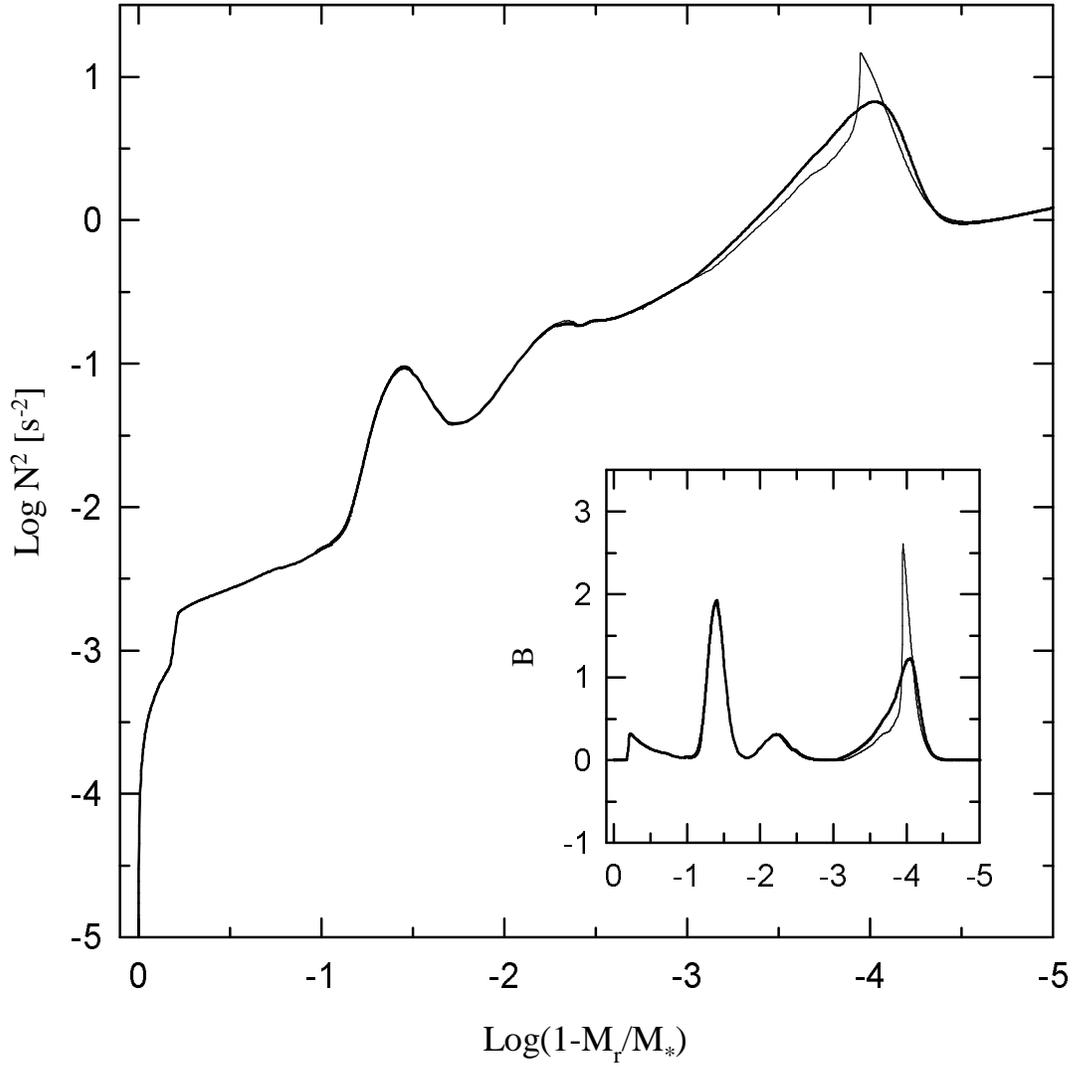}
\caption{The logarithm of the squared Brunt-V\"ais\"al\"a
frequency and the Ledoux  term, $B$, for the non-equilibrium diffusion
model.  The results for the diffusive equilibrium approximation are 
shown in thin lines.}
\end{figure*}

\begin{figure*}
\centering
\includegraphics[width=500pt]{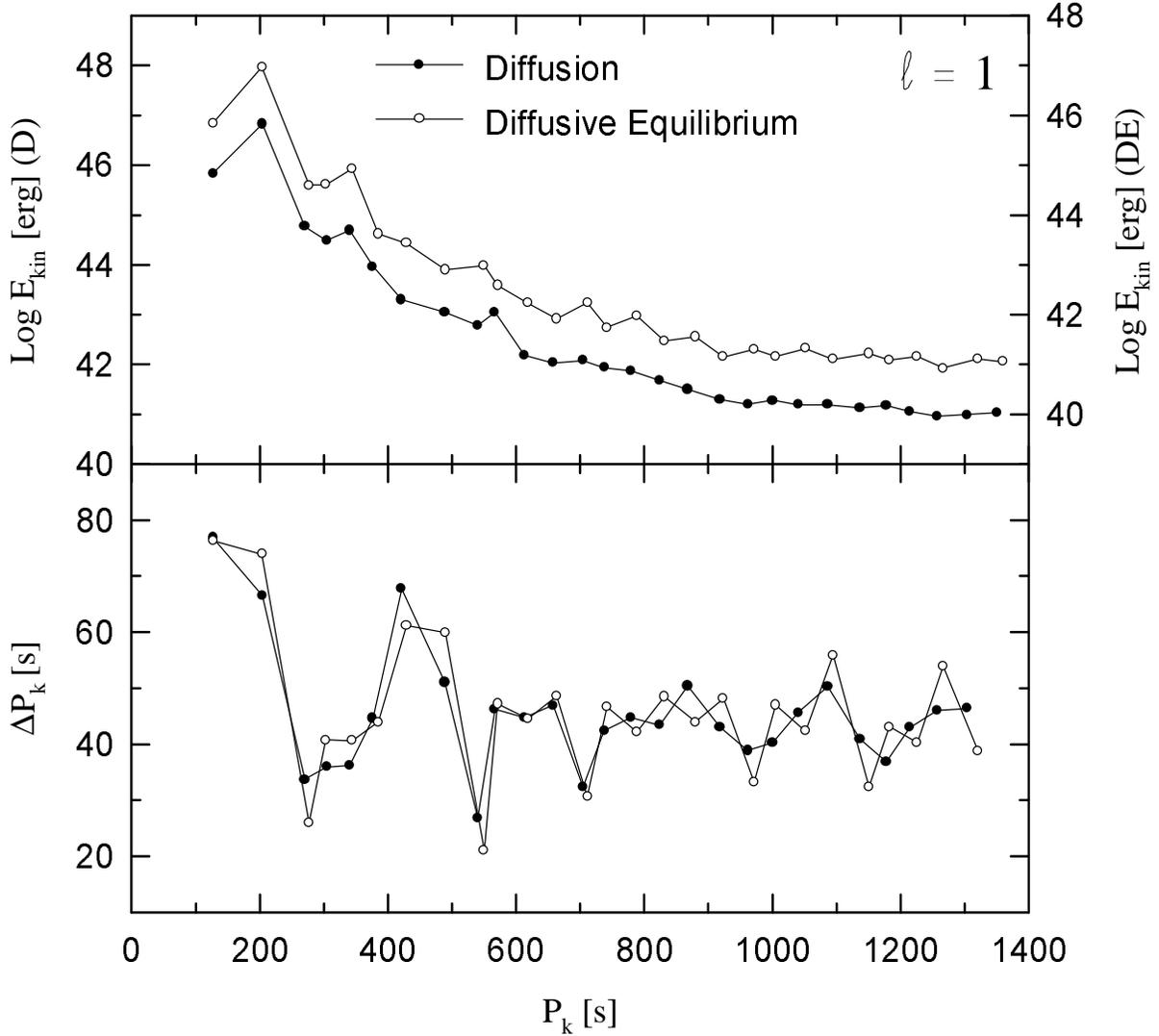}
\caption{Oscillation kinetic energy (upper panel) and period spacing 
(lower panel)  
values for $\ell=  1$ in terms of the  computed periods, $P_k$. Filled
dots  correspond to pulsational  computations for  the non-equilibrium
diffusion model, and empty dots for the diffusive equilibrium one.  In
the interests of clarity, the scale for the kinetic energy in the case
of  diffusive  equilibrium  is  displaced  upwards by 1 dex.   The
kinetic energy values correspond to the normalization $y_1= \delta r /
r = 1$ at the surface of the non-perturbed models for each mode.}
\end{figure*}

\begin{figure*}
\centering
\includegraphics[width=500pt]{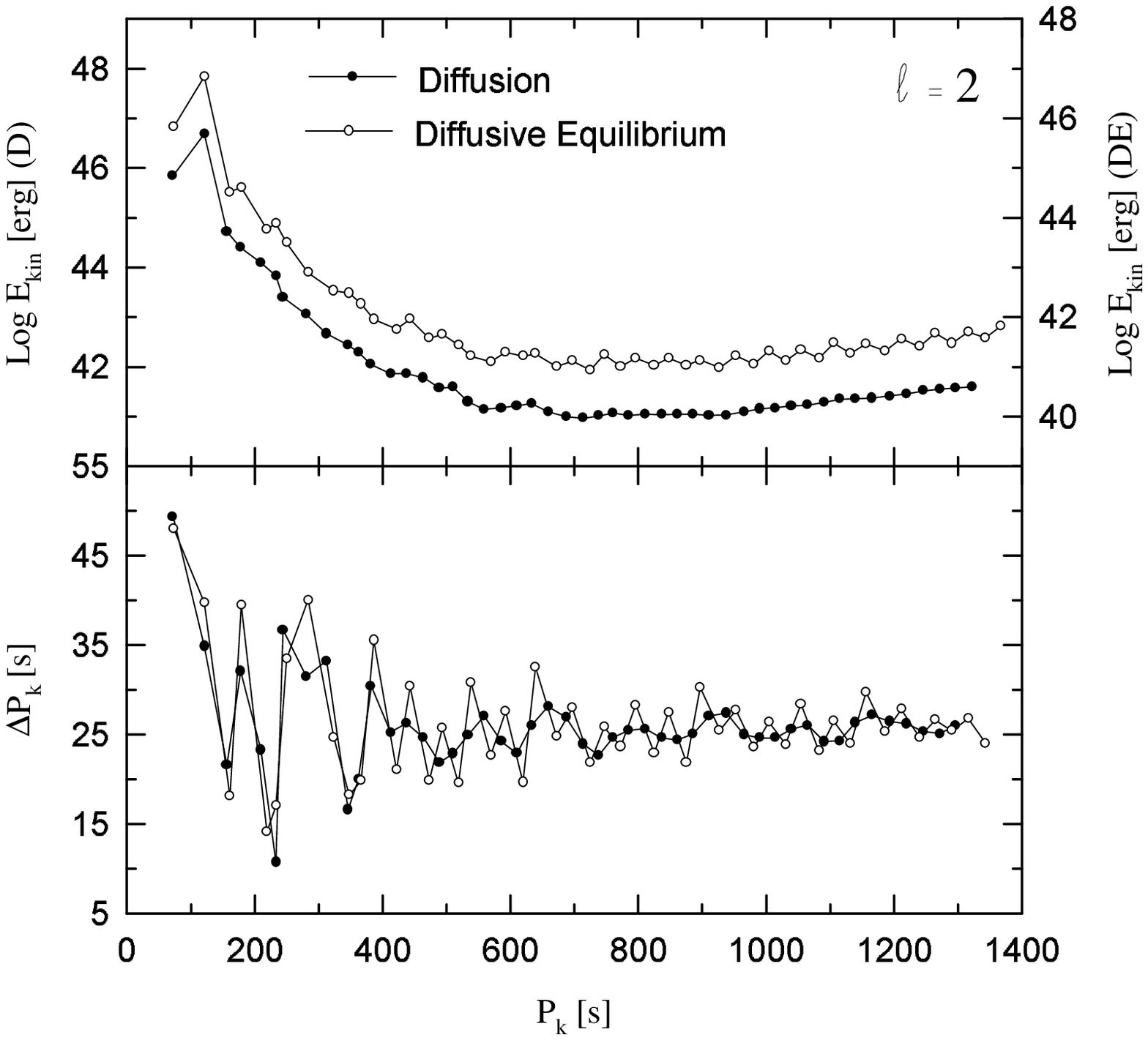}
\caption{Same as Fig. 3 but for $\ell= 2$.}
\end{figure*}

\begin{figure*}
\centering
\includegraphics[width=500pt]{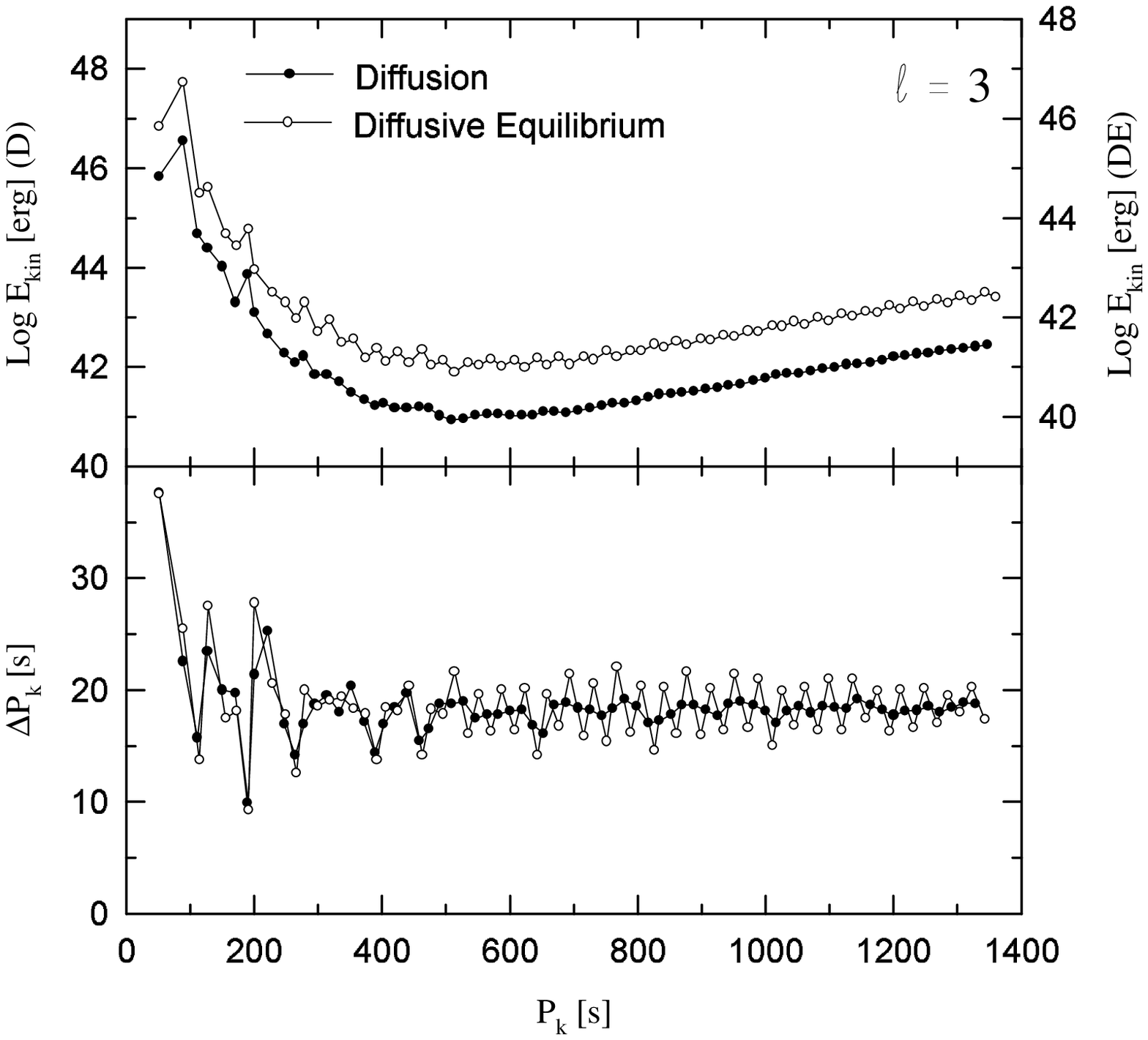}
\caption{Same as Fig. 3 but for $\ell= 3$.}
\end{figure*}

\end{document}